\documentclass[bibyear]{aa}  

\usepackage{graphicx}
\usepackage[varg]{txfonts}
\usepackage{url}
\usepackage[pdfborder={0 0 0 0},urlcolor=blue,breaklinks]{hyperref}
\hypersetup{colorlinks, citecolor=blue, filecolor=black, linkcolor=black, urlcolor=blue}
\usepackage[normalem]{ulem}	
\usepackage{natbib}
\usepackage{color}
\bibpunct{(}{)}{;}{a}{}{,}

\makeatletter

\newcommand{\Rmnum}[1]{\expandafter\@slowromancap\romannumeral #1@}
\makeatother
\newcommand{\rsun}{R$_{\odot}$}
\newcommand{\eg}{\emph{e.g.}}
\newcommand{\cf}{\emph{cf}}
\newcommand{\ie}{\emph{i.e.}}

\newcommand{\corr}[1]{\textcolor{black}{#1}}

\begin{document} 

\title{Measuring the magnetic field of a trans-equatorial loop system using coronal seismology}

\author{D.~M.~Long\inst{1} \and G.~Valori\inst{1} \and D.~P\'{e}rez--Su\'{a}rez\inst{1} \and R.~J.~Morton\inst{2} 
\and A.~M.~V\'{a}squez\inst{3,4}}

\institute{UCL--Mullard Space Science Laboratory, Holmbury House, Holmbury St. Mary, Dorking, Surrey, RH5 6NT, U.K. \\
\email{david.long@ucl.ac.uk}
\and Mathematics, Physics and Electrical Engineering, Northumbria University, Newcastle Upon Tyne, NE1 8ST, UK
\and Instituto de Astronom\'{i}a y F\'{i}sica del Espacio (IAFE), CONICET-UBA, Buenos Aires, Argentina
\and Facultad de Ciencias Exactas y Naturales (FCEN), UBA, Buenos Aires, Argentina}

\date{Received Jan 10, 2016; accepted Jan. 11, 2016}

\abstract
{``EIT waves'' are \corr{freely-propagating global} pulses in the low corona which are strongly associated with the initial 
evolution of coronal mass ejections (CMEs). They are thought to be \corr{large--amplitude, fast--mode magnetohydrodynamic} 
waves initially driven by the rapid expansion of a CME in the low corona.}
{An ``EIT wave'' was observed on 6~July~2012 to impact an adjacent trans--equatorial loop system which then 
exhibited a decaying oscillation as it returned to rest. Observations of the loop oscillations were used to estimate 
the magnetic field strength of the loop system by studying the decaying oscillation of the loop, measuring the 
propagation of ubiquitous transverse waves in the loop and extrapolating the magnetic field from observed magnetograms.}
{Observations from the Atmospheric Imaging Assembly onboard the \emph{Solar Dynamics Observatory} (\emph{SDO}/AIA) 
and the Coronal Multi-channel Polarimeter (CoMP) were used to study the event. An Empirical Mode Decomposition analysis 
was used to characterise the oscillation of the loop system in CoMP Doppler velocity and line width and in AIA intensity.}
{The loop system was shown to oscillate in the 2nd harmonic mode rather than at the fundamental frequency, with the 
seismological analysis returning an estimated magnetic field strength of \corr{$\approx5.5\pm1.5$~G}. This compares to the magnetic 
field strength estimates of $\approx$1--9~G and $\approx$3--9~G found using the measurements of transverse wave 
propagation and magnetic field extrapolation respectively.}
{}


\maketitle

\section{Introduction}\label{s:intro}

First observed by the \emph{Solar and Heliospheric Observatory} \citep[\emph{SOHO};][]{Domingo:1995}, many theories 
have been proposed to interpret globally--propagating coronal disturbances (commonly called ``EIT waves''). Taking 
their name from the Extreme ultraviolet Imaging Telescope \citep[EIT;][]{Dela:1995} onboard SOHO, they are typically 
observed as radially expanding bright features associated with the onset of a coronal mass ejection (CME) that can 
traverse the solar disk in under an hour \citep[\eg,][]{Moses:1997,Dere:1997,Thompson:1998}. After almost 20~years 
of detailed investigation and debate, a consensus is finally being reached with regard to the physics underpinning 
their evolution, thanks to the improved temporal and spatial resolution of the \emph{Solar and Terrestrial Relations 
Observatory} \citep[\emph{STEREO};][]{Kaiser:2008} and more recently the \emph{Solar Dynamics Observatory} 
\citep[\emph{SDO};][]{Pesnell:2012}.

The multitude of theories proposed to explain this phenomenon is mainly the result of conflicting observations 
\citep[\cf.][]{Long:2017}. Initially interpreted as the coronal counterpart of the chromospheric Moreton--Ramsey wave 
\citep{Moreton:1960a,Moreton:1960b}, ``EIT waves'' were treated as fast--mode magneto--acoustic (MHD) waves following 
the example of \citet{Uchida:1968}. However, issues with this interpretation were raised by observations of stationary 
brightenings at the edges of coronal holes \citep[\cf.][]{Delannee:1999}. These observations led to the suggestion 
that ``EIT waves'' were not true ``waves'' but instead were a brightening produced by the restructuring of the magnetic 
field during the eruption of a CME. It was proposed that this brightening was alternatively produced by Joule heating 
at the interface between the magnetic field of the erupting CME and the surrounding coronal magnetic field 
\citep{Delannee:2000,Delannee:2008}, continuous reconnection of small--scale magnetic loops driven by the erupting CME 
\citep{Attrill:2007} or the stretching of magnetic field lines overlying an erupting flux rope \citep{Chen:2002,Chen:2005}. 
This last scenario was also supported by the relatively low observed speed of the disturbances \citep[with an average 
speed of 200--400~km~s$^{-1}$ measured by][using observations from \emph{SOHO}/EIT]{Thompson:2009}.

However, these hypotheses have been undermined both by observations of reflection and refraction at coronal hole and 
active region boundaries \citep[\eg,][]{Thompson:2000,Long:2008,Veronig:2008,Gopal:2009,Shen:2013,Kienreich:2013} as 
well as the higher speeds measured using \emph{STEREO} \citep[100--630~km~s$^{-1}$;][]{Muhr:2014} and \emph{SDO}
\citep[v$_{mean}\approx644$~km~s$^{-1}$;][]{Nitta:2013}. Although initially analysed using the linearised fast--mode wave 
equations, observations of pulse dispersion and deceleration \citep{Warmuth:2004a,Warmuth:2004b,Long:2011a,Long:2011b} have 
led to their interpretation as large--amplitude simple waves \citep[\eg,][]{Vrsnak:2000a,Vrsnak:2000b,Warmuth:2007,Vrsnak:2008} 
initially driven by the rapid expansion of the erupting CME in the low corona \mbox{\citep{Patsourakos:2010}} before 
propagating freely. Note that a number of different reviews by \citet{Gallagher:2011}, \citet{Zhukov:2011}, 
\citet{Patsourakos:2012}, \citet{Liu:2014} and more recently \citet{Warmuth:2015} discuss the different interpretations 
and the observations both supporting and contradicting them in detail.

More recently, the higher temporal and spatial resolution provided by the Atmospheric Imaging Assembly \citep[AIA;][]{Lemen:2012} 
has revolutionised our understanding of ``EIT waves''. These improved observations are providing clear evidence that the freely 
propagating ``EIT waves'' behave as waves, in principle allowing their observed characteristics to be used as diagnostics of 
the physical properties of the coronal regions through which they propagate \citep[an approach called ``coronal seismology'', 
\eg,][]{Roberts:1984,Ballai:2007} and estimate properties such as magnetic field strength 
\corr{\citep[\cf.][]{Warmuth:2005,West:2011,Kwon:2013,Long:2013}}. A similar approach may also be applied on a more local scale, 
using the forced oscillations of coronal loops initially driven by the impact of an ``EIT wave'' to determine properties such as 
their magnetic field strength or the energy of the wave-pulse \citep[\eg,][]{Ballai:2007,Ballai:2008,Ballai:2011,Yang:2013}. 

While the global nature of ``EIT waves'' greatly increases the chances of them interacting with coronal loops, these 
observations are dependent on a sufficiently high temporal and spatial resolution to be able to identify the oscillating 
loop. Despite its global field-of-view, \emph{SOHO}/EIT did not have a sufficiently high temporal or spatial resolution 
to identify oscillating loops. This changed with the launch of the \emph{Transition Region And Coronal Explorer} 
\citep[\emph{TRACE};][]{Handy:1999}, whose observations were used to show that the oscillation of a coronal loop may 
be used to estimate its magnetic field strength \citep[\eg,][]{Aschwanden:1999,Naka:2001,Aschwanden:2011,Guo:2015}. 
Subsequent work extended the method to observations of oscillation in Doppler motion made by the Extreme ultraviolet 
Imaging Spectrometer \citep[EIS;][]{Culhane:2007} onboard the \emph{Hinode} spacecraft \citep[\eg,][]{Vandoor:2008a}.

In this paper, we estimate the magnetic field of a trans--equatorial loop system using multiple independent techniques. The 
first approach uses the oscillation of the loop system resulting from the impact of the global EUV wave-pulse, the second uses 
direct observations of the magnetic field made by the by the ground–based Coronal Multi–channel Polarimeter (CoMP) instrument
\citep{Tomczyk:2008}, while the third estimates the field strength using two independent magnetic field extrapolations from 
the photosphere. \corr{This paper builds on work previously presented at the International Astronomical Union (IAU) Symposium 
on ``Solar and Stellar Flares and Their Effects on Planets'' \citep[described by][]{Long:2016}, and represent a significant 
advance on the previously published proceedings paper.} The observations are presented in Section~\ref{s:obs}, with the 
properties of the pulse and its interaction with the surrounding corona examined in Sections~\ref{s:pulse} and \ref{s:impact} 
respectively. This interaction is then used to derive the magnetic field strength associated with a nearby transequatorial 
loop system in Section~\ref{s:bfield}. Finally, some conclusions about the implication of these observations are drawn in 
Section~\ref{s:conc}.

\section{Observations \& Data Analysis}\label{s:obs}

\begin{figure}[!t]
\centering{
               \includegraphics[width=0.49\textwidth,clip=,trim=0mm 0mm 0mm 0mm]{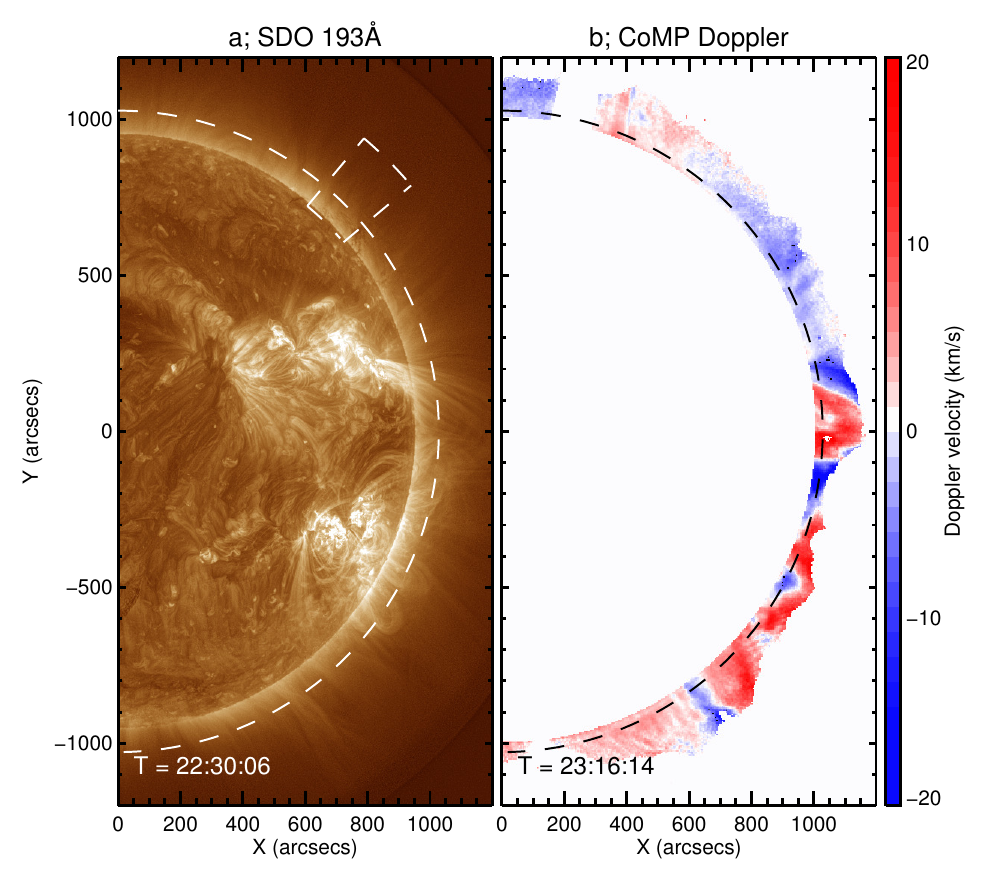}
               }
\caption{Intensity image from \emph{SDO} 193~\AA\ (panel~a) and Doppler velocity from CoMP (panel~b). The dashed lines above the 
         limb (at a height $1.09\,{\rm R}_\odot$) in both panels are used to make Figure~\ref{fig:193_aia_comp}. \corr{The arc 
         sector in the top right of panel~a shows the region used to get the density of the quiet Sun in Section~\ref{ss:seis}.} 
         Note that the intensity image in panel~a has been enhanced using the Multi--scale Gaussian Normalisation technique of 
         \citet{Morgan:2014}. \corr{The evolution of the 193~\AA\ channel is shown in the left panel of the movie available 
         online.}}
\label{fig:context}
\end{figure}

\begin{figure*}[!ht]
\centering{
           \includegraphics[width=0.9\textwidth,clip=,trim=0mm 0mm 0mm 0mm]{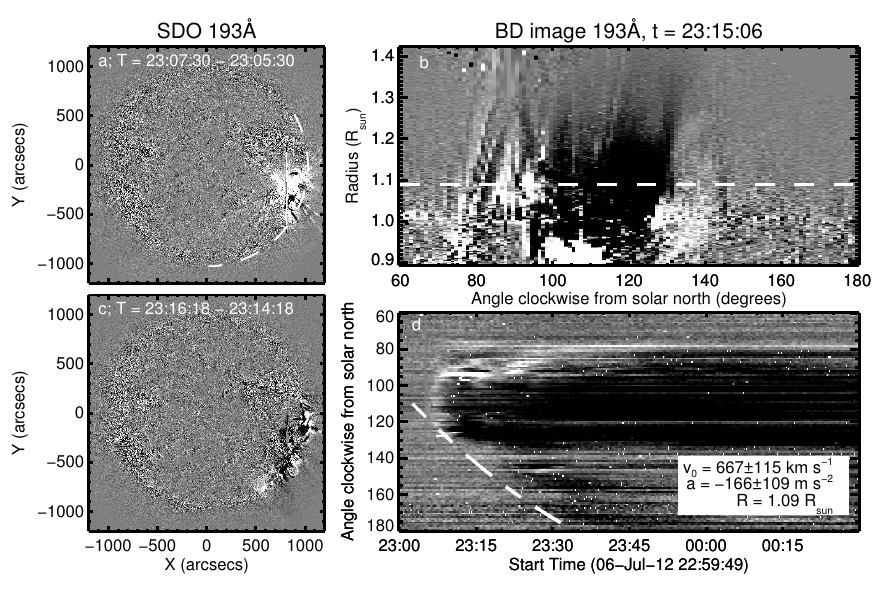}
           }
\caption{Running difference images from the 193~\AA\ passband of the eruption (panel~a) and the evolution of the wave 
         (panel~c). \corr{The temporal evolution of the wave is shown in the right panel of the movie available online 
         (showing the west half of the solar disk only).} Panel~b shows a base--difference deprojected annulus image 
         showing the eruption at 23:15:06~UT, with the vertical axis giving the height from Sun-centre and the horizontal 
         axis giving the position angle clockwise from solar north (indicated by the dashed white line in panel~a). 
         Panel~d shows a base--difference image of the temporal variation at a height of 1.09~\rsun\ (indicated by the 
         white dashed line in panel~b). The dashed white line in panel~d shows the fit to the leading edge of the 
         propagating pulse.}
\label{fig:eruption}
\end{figure*}

The solar eruption studied here originated from NOAA active region AR~11514 on 6~July~2012 and was associated with a CME 
and a GOES X1.1 class flare which began at 23:01~UT. The event was well observed by multiple instruments including 
\emph{SDO}/AIA and CoMP (see Figure~\ref{fig:context}), providing an opportunity to study the eruption in detail. As with 
the event of 25~February~2014 previously studied by \citet{Long:2015}, the global EUV wave observed here did not propagate 
isotropically, (see Figure~\ref{fig:eruption}). Instead, due to the presence of the adjacent active regions AR~11515 to the 
East and ARs~11513, 11516 and 11517 to the North, the ``EIT wave'' propagated mainly towards the south polar coronal hole 
along the limb as seen by \emph{SDO}/AIA (this \corr{is shown in the movie attached to Figure~\ref{fig:context}}). 

Designed to study the coronal magnetic field, CoMP provides Stokes\emph{-I} measurements in the \ion{Fe}{xiii} 
10747~\AA\ and 10798~\AA\ emission lines with a field of view of 2.8~\rsun\ and an image size of $620\times620$ 
pixels. This gives an image sample size of $\sim$4.25~arcsec~pixel$^{-1}$ at $\sim$30~s cadence \citep[\cf.][]{Tian:2013}. 
Although the seeing was not good enough for this event to estimate the full Stokes parameters (S.~Tomczyk, private 
communication), the Stokes\emph{-I} measurements can be fitted using a least--squares Gaussian fit to estimate the line 
intensity, width and central wavelength for each pixel. This allows the plasma parameters to be studied, giving an estimate 
of the temporal variations in Doppler motion and line width in the low corona.

\section{Pulse properties}\label{s:pulse}

As the “EIT wave” was observed to propagate along the solar limb from the erupting active region towards the south pole, it was 
not possible to use the \textsf{Coronal Pulse Identification and Tracking Algorithm} \citep[\textsf{CorPITA;}][]{Long:2014} to 
study the propagation of the pulse. Instead, following \citet{Long:2015}, a polar deprojection was used to determine the variation 
in pulse kinematics across a height range from 1.01--1.12~\rsun\ (as shown in Figure~\ref{fig:eruption}). This 
height range was chosen because above $\approx$1.12~\rsun\ the pulse becomes very faint in the \emph{SDO}/AIA images while the 
CoMP observations become noisy and prone to missing data, making direct comparisons between the instruments difficult. 

The leading edge of the wave--pulse was manually identified and fitted using a quadratic model at each height across the 
entire range, as shown for height 1.09~\rsun\ in Figure~\ref{fig:eruption}, with the process repeated 10 times in each case to 
minimise uncertainty. The pulse was found to have a velocity ranging from 607--1583~km~s$^{-1}$, with a mean velocity of 
$\approx$1106$\pm314$~km~s$^{-1}$ and an acceleration ranging from $-376$ -- $-19$~m~s$^{-2}$, with a mean acceleration of 
$\approx-207\pm107$~m~s$^{-2}$. These estimates are much higher than the average ``EIT wave'' speed measured by 
\citet{Nitta:2013}, indicating that the pulse measured here was quite fast. The pulse also exhibited clear deceleration, 
evidence of broadening and was associated with a Type~\Rmnum{2} radio burst \mbox{\citep[\cf.][]{Long:2017}}, suggesting that 
it was a large amplitude wave pulse.

For such a fast and intense pulse it is possible to follow the approach of \citet{Long:2015} and estimate its initial energy 
using the Sedov--Taylor approximation \citep{Taylor:1950a,Taylor:1950b,Sedov:1959}. Although this assumes a spherically 
symmetric blast wave emanating from a point source (which is not strictly valid here), \citet{Grechnev:2008,Grechnev:2011a} 
and \citet{Long:2015} found that the approach is suitable for analysing the onset stage of pulses being initially driven over 
a very short time period before propagating freely, as with the event studied here. \citet{Long:2015} have also shown that the 
Sedov--Taylor relation provides an excellent estimate of the initial energy of the eruption assuming a blast wave propagating 
through a medium of variable density. Using this approach, the initial energy of the pulse was estimated to be 
$\sim$8.6$\times$10$^{31}$~ergs, comparable to the previous estimate made by \citet{Long:2015}\corr{. While this is consistent
with the observations of high initial velocity and strong pulse deceleration, it is most likely an overestimate of the true energy
of the wave-pulse. As noted by \citet{Long:2015}, the Sedov-Taylor relation assumes a spherical blast wave emanating from a source
point, which is not the case here and does not include the effect of the coronal magnetic field. As a result, this estimate should 
be considered as a first order approximation of the energy of the pulse.} 

\begin{figure*}[t]
\centering{
               \includegraphics[width=0.99\textwidth,clip=,trim=5mm 5mm 0mm 0mm]{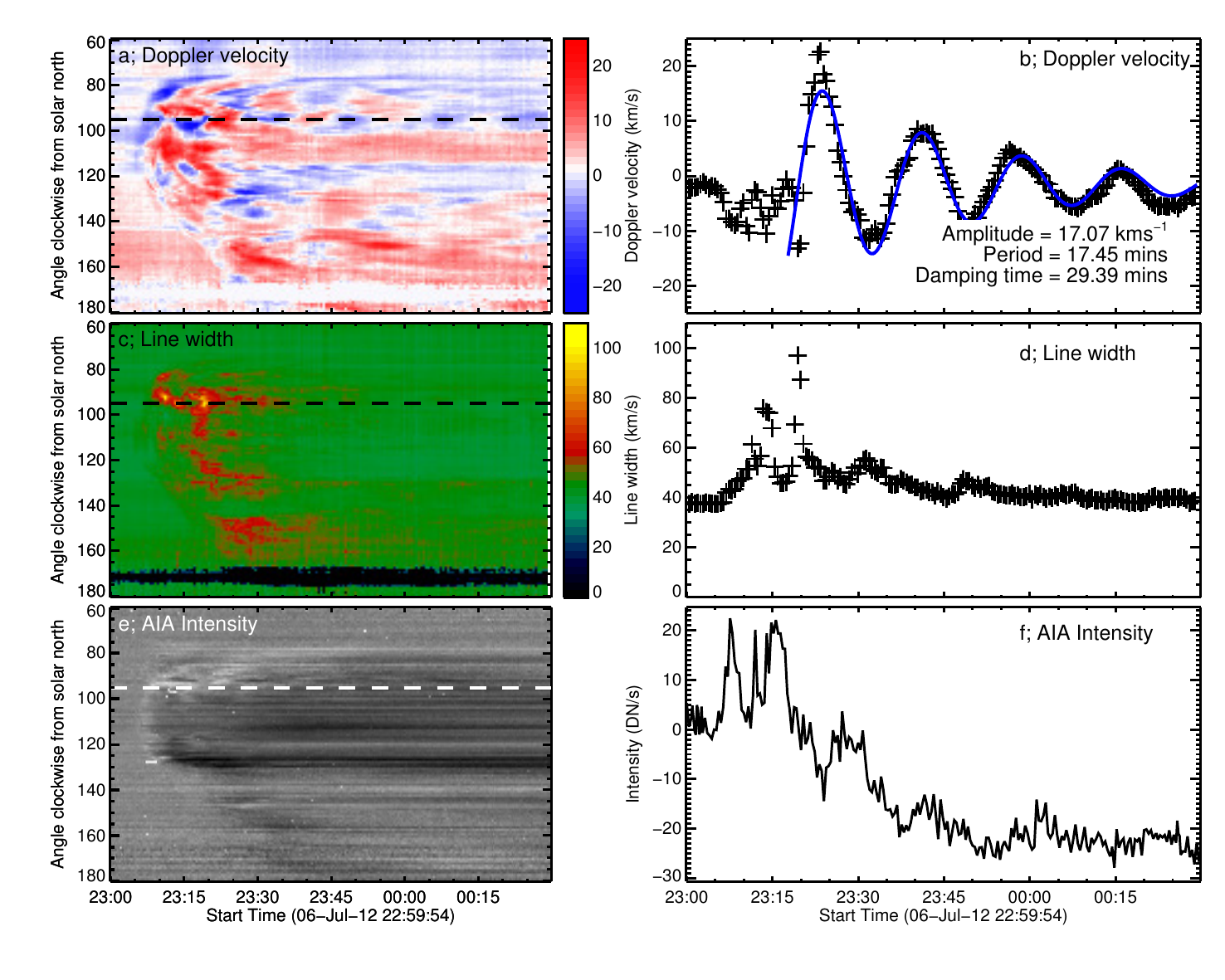}
               }
\caption{Temporal variation in CoMP Doppler velocity (panel~a) and line width (panel~c) and AIA 193~\AA\ percentage 
	      base difference (PBD) intensity (panel~e) at a height of 1.09~\rsun\ (shown by the dashed line in panels~a \& b 
	      of Figure~\ref{fig:context}). Panels~b, d \& f show the temporal variation in Doppler velocity, line width and AIA 
	      PBD intensity along the dashed line indicated in panels~a, c \& e. The oscillatory behaviour of the loop system is 
          very clear in the Doppler motion (panels~a \& b), indicating motion to-and-from the observer, while the impact of 
          the pulse and subsequent evacuation of plasma into the CME is apparent in the line width (panels~c \& d) and 
          plane-of-sky AIA observations respectively (panels~e \& f), as described in the text. The solid blue line in panel~b 
          indicates a fit to the data using the damped cosine model described in Equation~\ref{eqn:cos}.}
\label{fig:193_aia_comp}
\end{figure*}

Although the pulse can be clearly identified in the AIA 193~\AA\ \corr{intensity} observations (\eg, Figure~\ref{fig:eruption}), 
it was not as apparent in the CoMP \corr{intensity} observations. This is clear from panel~a of Figure~\ref{fig:193_aia_comp}, 
where the effects of the pulse can be seen in Doppler velocity, but not the pulse itself. As a result, the CoMP observations were 
used to study the effects of the pulse on the surrounding corona, with the AIA 193~\AA\ observations used to estimate the pulse 
kinematics.

\section{Interaction with trans--equatorial loop system}\label{s:impact}

Although the Sedov--Taylor approximation assumes an isotropic expansion of the wave--pulse being studied, this is not the case 
here due to the trans--equatorial loop system to the north of the erupting active region. This is shown in Figure~\ref{fig:eruption} 
and the associated movie movie1.mov. While this feature restricts the propagation of the wave--pulse, the effects of the impact 
force a significant displacement of the trans--equatorial loop system from rest. This results in a large amplitude decaying 
oscillation in CoMP Doppler observations as the loop returns to its pre-impact position, as shown in detail in 
Figure~\ref{fig:193_aia_comp} at a specific location of the observed loop structure. As a result, it is possible to estimate the 
magnetic field of the loop system using a coronal seismology approach \citep[\eg,][]{Roberts:1984,Vandoor:2008a,Long:2013}.

The temporal variation in CoMP Doppler velocity and line width and AIA percentage base difference (PBD) intensity at a height 
of 1.09~\rsun\ and for all polar angles are shown in the left panels (a,c,e) of Figure~\ref{fig:193_aia_comp}. The pulse is 
clearest in the Doppler velocity and AIA PBD measurements (panels a \& e), although there is a slight suggestion of variation 
in the line width measurements (panel c). The wave--pulse is first seen in the Doppler velocity observations at $\approx$23:04~UT, 
with a slightly blue--shifted edge moving northwards away from the erupting active region (located at $\sim$110$^{\circ}$ 
clockwise from solar north). It can also \corr{be} seen to move towards the south pole at roughly the same time, again observed as an 
initially slightly blue--shifted edge. A co-temporal faint bright feature can also be identified in the PBD measurements in panel~e 
moving both north and south away from the source active region. Although there is some indication of a slight change in the line 
width shown in panel~c at this time, there is no clear signature of a propagating front. 

\begin{figure*}[!ht]
\centering{
               \includegraphics[width=0.985\textwidth,clip=,trim=5mm 2mm 0mm 0mm]{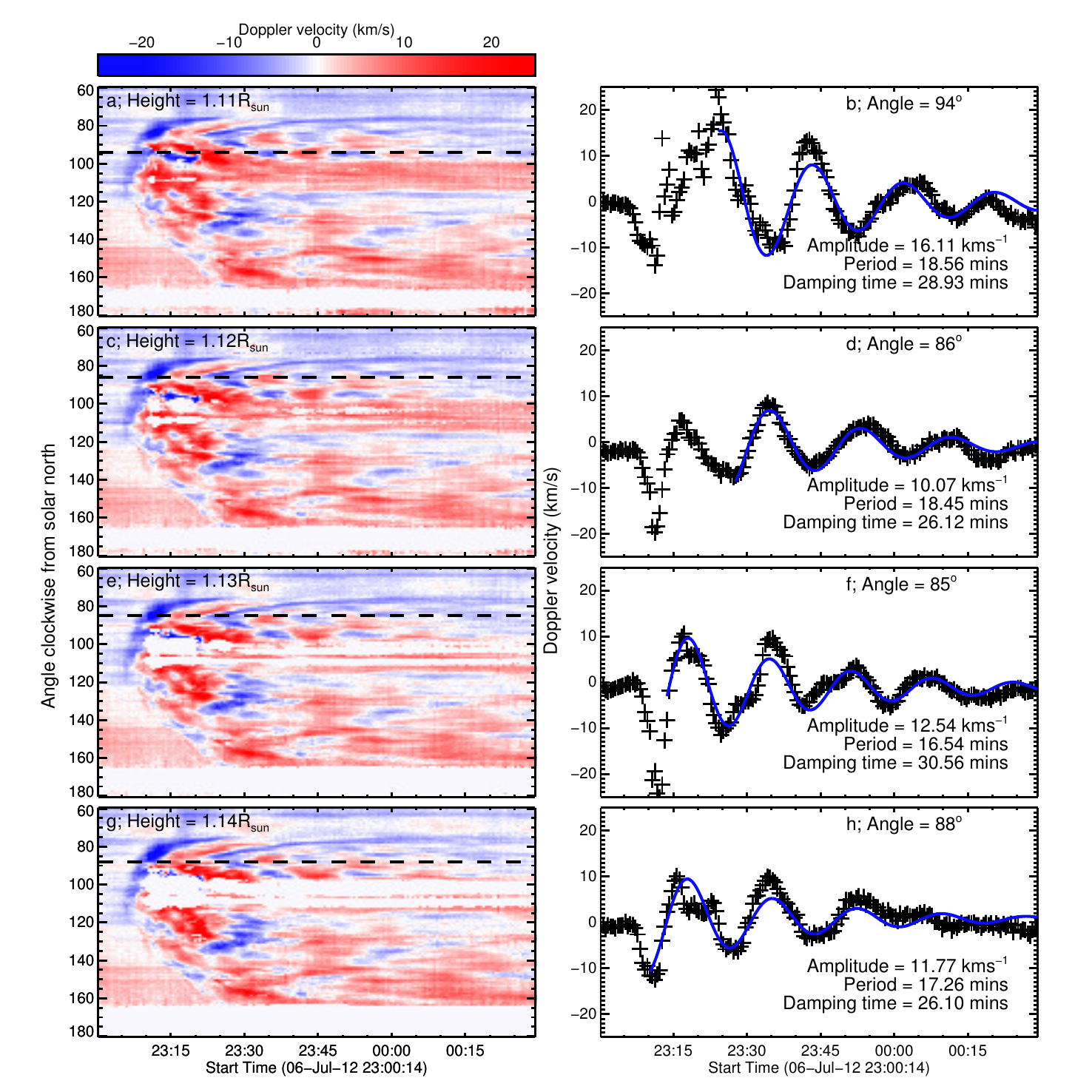}
               }
\caption{Temporal variation in Doppler velocity at heights of 1.11, 1.12, 1.13 \& 1.14~\rsun\ for all polar angles (left column) 
		 along with its temporal profile measured at angles of 94$^{\circ}$, 86$^{\circ}$, 85$^{\circ}$ \& 88$^{\circ}$ clockwise 
         from solar north (right column). The solid blue lines in the right column indicate a fit to the data using the damped cosine 
         model described in Equation~\ref{eqn:cos}.}
\label{fig:comp_dopp_profs}
\end{figure*}

Following the initial front, there is clear evidence of material being ejected with the erupting CME. This is apparent from the 
red--shifted outflow from the erupting active region apparent in the sector from 100--120$^{\circ}$ starting at $\sim$23:07~UT 
and continuing for the rest of the time period shown. This corresponds to a drop in the PBD intensity (panel~e of 
Figure~\ref{fig:193_aia_comp}), indicating a drop in density and/or temperature as material is evacuated by the CME. This 
drop in intensity is also clear in panel~f, which shows the temporal variation in PBD intensity at $\sim$95$^{\circ}$ clockwise 
from solar north (as for panels~b and d, shown by the dashed line in panel~e). 

\begin{figure*}[!t]
\centering{               
	\includegraphics[width=0.9\textwidth,clip=,trim=0mm 0mm 0mm 0mm]{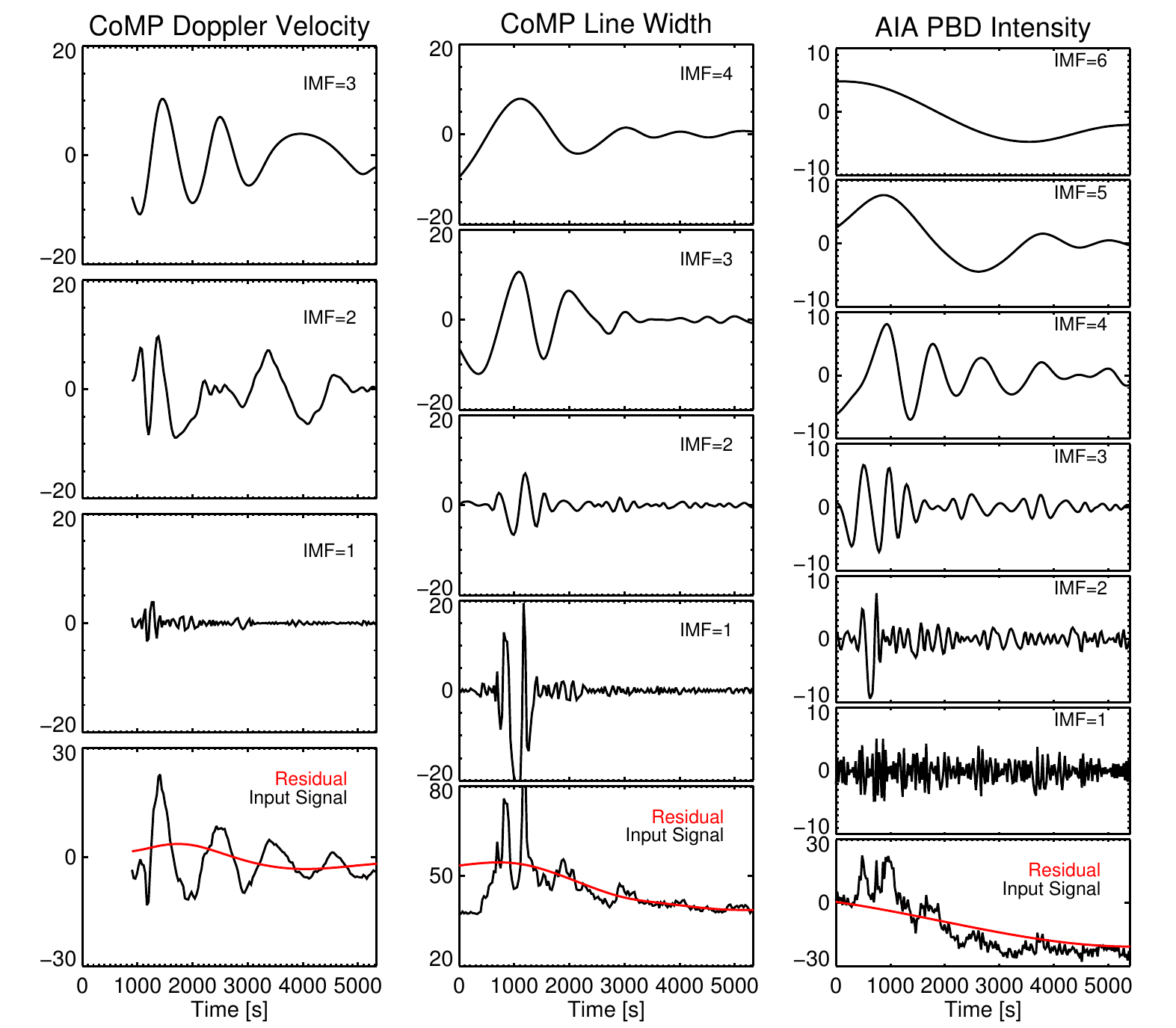}
    }
\caption{Intrinsic Mode Functions derived from the time evolution of the Doppler velocity (left panels), line width (centre panels) 
		 and AIA PBD intensity (right panels), corresponding to the data in the right panels of Figure~\ref{fig:193_aia_comp}. In 
         each case the bottom panel shows the original signal (\cf. Figure~\ref{fig:193_aia_comp}b, d \& f) with the upper panels 
         giving the derived components.}
\label{fig:imf_output}
\end{figure*}

While the propagation of the pulse to the south of the erupting active region is relatively uninhibited, the propagation to the 
north is modified by a trans--equatorial loop system located between $\sim$85--100$^{\circ}$ clockwise from solar north (as 
shown in Figure~\ref{fig:context}). The variation in Doppler velocity with time in panel~a of Figure~\ref{fig:193_aia_comp} 
shows that this loop system is relatively stable until the impact of the blue--shifted wave--pulse at $\sim$23:05~UT. A more 
strongly blue--shifted feature is then observed between $\sim$23:07~UT and $\sim$23:17~UT along the profile in panel~b, 
which is also characterised by a very strong co-temporal increase in line width (seen in panel~d). This suggests turbulent 
behaviour \citep[\cf.][]{Harra:2009}, and is consistent with the initial impact of the wave--pulse on the loop system followed by 
the outward motion of the associated erupting filament. 

The effects of the impact of the wave--pulse on the the trans--equatorial loop system may then be seen in panel~a from 
$\sim$23:17~UT onwards as it exhibits a series of alternating red-- and blue--shifted features. This can be attributed to the 
loop system being displaced from its rest position by the impact of the wave--pulse and subsequently returning to rest via a 
decaying oscillation. This behaviour is shown most clearly in panel~b of Figure~\ref{fig:193_aia_comp}, which corresponds 
to the temporal Doppler velocity profile at an angle of $95^\circ$ clockwise from solar north (indicated by the dashed line in 
panel~a). It should also be noted that as the Doppler velocity in panel~b begins to exhibit this strong oscillation, the line width 
drops dramatically to near the pre-event level, suggesting a near-uniform oscillation of the loop system.

This oscillatory behaviour in the Doppler velocity is consistent along the loop as shown in Figure~\ref{fig:comp_dopp_profs}, which 
shows the variation in Doppler velocity at a set of sampled locations (height and polar angle) along the rest of the loop system. 
It is clear from the temporal evolution of the Doppler velocities shown in the left panels of Figure~\ref{fig:comp_dopp_profs} 
that the signal weakens above $\sim$1.14~\rsun, with clear data drop-outs apparent at $\sim$95--110$^{\circ}$. While the 
amplitude of the oscillation can be seen to drop with increasing height, the period and damping time are comparable in all cases, 
suggesting that the values quoted in Figure~\ref{fig:193_aia_comp} are representative of the oscillation along the loop. These 
observations suggest that the wave-pulse impacts the leg of the loop system in the low corona (thus leading to the larger amplitude 
of the oscillation at lower heights). The consistent damping time also suggests that while the loop system is perturbed by the 
pulse impact, it stays relatively stable and is not opened by this impact. The mass of the loop system and hence the oscillation 
and associated damping coefficient therefore remain constant throughout the oscillation.


\subsection{Empirical Mode Decomposition analysis}\label{ss:emd}

\begin{figure*}[!t]
\centering{
	\includegraphics[width=0.98\textwidth,clip=,trim=0mm 0mm 0mm 0mm]{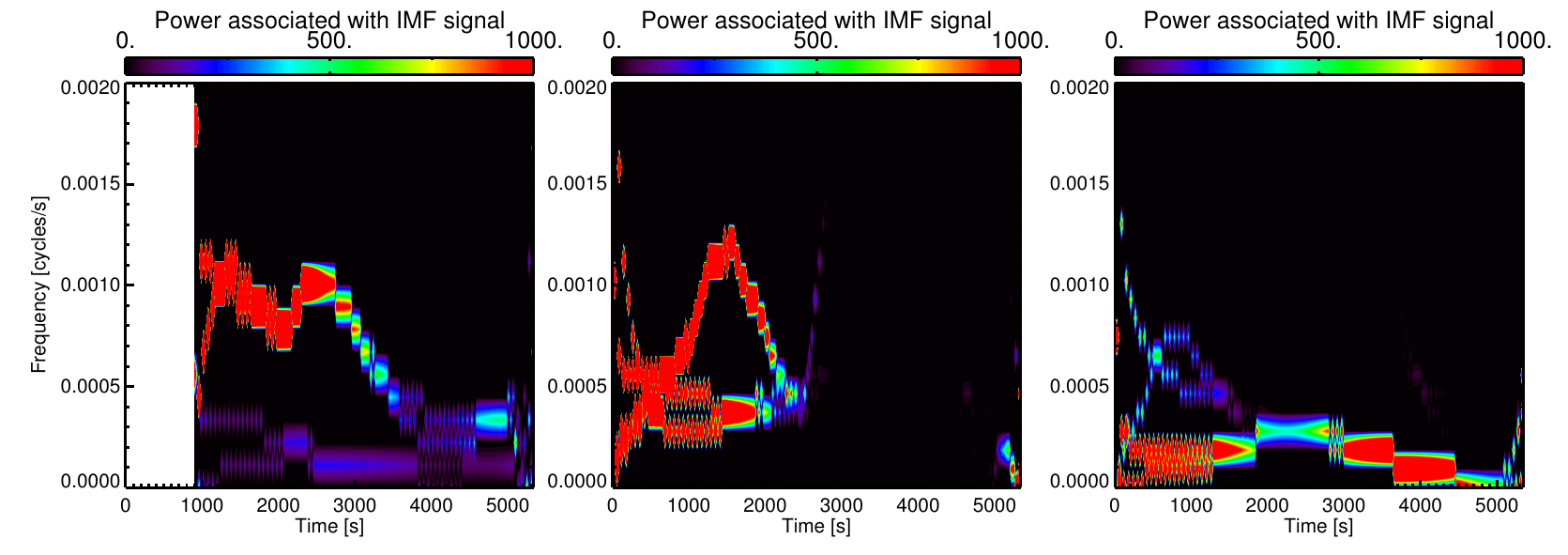}
    }
\caption{Hilbert Huang Transform \corr{(HHT)} spectra for the Doppler velocity (left column), line width (centre column) 
         and PBD intensity (right column) corresponding to the IMFs shown in Figure~\ref{fig:imf_output}.
          \corr{Note that only the HHT of the IMF=3 and residual, IMF=3 and 4, and IMF=4 and 5 are shown in the the three 
          cases, respectively.}
          }
\label{fig:hht_output}
\end{figure*}

While the oscillatory behaviour of the Doppler velocity is very clear in panel~b of Figure~\ref{fig:193_aia_comp}, there is some 
indication of a comparable (albeit \corr{extremely} faint) oscillation in the line width and PBD intensity shown in panels~d \& f. 
To determine if an oscillation was present, the temporal evolution of the different signals was examined using an Empirical Mode 
Decomposition \citep[EMD;][]{Huang:1998,Quek:2003} analysis. This technique decomposes the original signal into a series of 
Intrinsic Mode Functions (IMFs) and a residual, with the associated energy distribution represented by a Hilbert-Huang spectrum. 
\corr{The approach involves constructing a low- and a high-envelope from the series of maxima and minima of the the signal, and 
then averaging the two envelopes. Under certain constraints, the resulting function is called an IMF and captures the fastest 
oscillating part in the signal. A new signal is then obtained by subtracting the IMF from the signal itself. This process is 
repeated until the updated signal shows no more oscillation, leaving the residual, \citep[see, \eg][for more details]{Stangalini:2014}. 
Since IMFs are time-dependent, a convenient way of representing their contribution to the signal at each time is to use an Hilbert 
transform as defined by \citet{Huang:1998}, which can be used to describe the time-evolution of the (instantaneous) frequency of each 
IMF.} 

The EMD analysis was applied to the Doppler velocity, line width and PBD intensity signals shown in 
panels~b, d \& f, respectively, of Figure~\ref{fig:193_aia_comp}. Note that in the following analysis we focus on the impact of 
the wave--pulse on the trans-equatorial loop system and its resulting oscillation, leaving a more detailed study of the correlation 
of the various IMFs between the different signals for a dedicated future work. 

Figure~\ref{fig:imf_output} shows the output from the EMD analysis, with the bottom panel for each column giving the original signal 
(black) and residual (red) while the upper panels give the corresponding IMFs. The left panels show that the Doppler velocity can be 
decomposed into three IMFs, with the oscillatory behavior mostly captured by IMF=3, which reveals an almost constant frequency of 
\corr{1.0~mHz up to about t=3000s}. 
While this approach can provide a more accurate estimate of the frequency of the oscillation, the composing IMFs are not orthogonal 
functions, leading to the crosstalk observed in the second half of the time series for IMF=2 \& 3. The IMF=2 contribution also shows 
the presence of a more complex component which is not purely oscillatory in nature. 
\corr{The residual also shows an oscillatory behavior on a longer period than the loop oscillation. However, the oscillation in the 
residual is not complete and therefore the residual is not considered to be an IMF.} The difference between the oscillatory Doppler 
signal and this 
\corr{longer} component is best seen using a Hilbert-Huang Transform (HHT) as shown in Figure~\ref{fig:hht_output}. The left panel here 
shows the HHT spectrum restricted to \corr{the residual and the IMF=3}. 
The spectral contribution of the 
\corr{oscillatory} component from IMF=
\corr{3} is represented by the intense red strip around 1~mHz which stays almost constant in frequency until $t\simeq3000$~s, before 
strongly damping and  decreasing to 0.5~mHz, \corr{(cf. the left panel of Figure~\ref{fig:imf_output}). The residual contribution appears 
in this plot as a} periodic, low-intensity component in the lower part of the spectrum at a frequency of 0.2~mHz. 

The central panels of Figure~\ref{fig:imf_output} show the decomposed IMF for the line width shown in panel~d of 
Figure~\ref{fig:193_aia_comp}. Here the high-frequency IMFs=1 \& 2 capture a transient nonlinear pulse in the time period 
$\sim600<t<1400$~s which has a period much shorter than the loop oscillation and cannot be clearly identified in the Doppler signal 
(except for a small-amplitude trace in the IMF=1 of the left-hand column of Figure~\ref{fig:imf_output}). \corr{The frequency of the 
IMF=2 in Figure~\ref{fig:imf_output} can be seen to chirp in time from $\sim$1.5 to 3.5~mHz, whereas the evolution of the IMF=1 is 
less clear. However for clarity neither IMF is included in Figure~\ref{fig:hht_output}. These signals are} clearly different to the 
loop system damped oscillation \corr{(mostly captured in Figure~\ref{fig:imf_output} by IMF=3), which again reproduces the component 
of the oscillating loop system in the IMF=3 of the Doppler velocity signal (although at a slightly more variable frequency between 
$\sim$0.5 and 1~mHz, see the central panel in Figure~\ref{fig:hht_output}), and a lower frequency component at about 0.4~mHz.}

The IMF decomposition of the AIA signal in the right panels of Figure~\ref{fig:imf_output} shows the highest level of complexity, 
partly due to the higher cadence of the AIA signal which captures more time-scales than CoMP. Despite this, both loop-system 
oscillation components can be identified with a clear damping signature (\eg, IMF=4), and the nonlinear signal due to the wave--pulse 
itself (\eg, IMF=2). \corr{In the HHT shown in the far right panel of Figure~\ref{fig:hht_output}, the frequency of the oscillating mode 
is similar to the one generated by the IMF=3 of the CoMP line-width signal, albeit relatively weaker. Similarly, the 0.2~mHz signal is 
also found in the PBD intensity from AIA.} The effects of the higher AIA cadence are seen in the highest-frequency IMF=1, which has the 
typical signature of noise and/or under-sampling (\ie, fluctuations on the highest frequency, whereas all higher IMFs are resolved signals).

The initial transients apparent in the IMF=1 of the CoMP line width and in the IMF=2 of the AIA PBD intensity are very similar 
in both structure and timing. The transients occur at the time the pulse reaches the trans-equatorial loop system, and appear 
as an oscillating wave modulated by an envelope rather than as a damped oscillation. It is intriguing that exactly such a pattern 
was previously assumed as the internal structure of an ``EIT wave'' pulse by \citet{Long:2011a} and we are therefore tempted to 
identify this transient signal as the pulse itself. 

However, if our pulse velocity estimate measured in Section~\ref{s:pulse} can also be applied at the front impacting the 
trans-equatorial loop system, this signal would indicate a pulse much broader than that observed. Therefore, we cannot 
unequivocally exclude that the observed transient signals are due to other nonlinear interactions between the pulse and loop 
systems. The case studied here is not optimal for discriminating between these possibilities and we leave such an analysis to a 
future work.

\section{Magnetic field strength estimates}\label{s:bfield}

\begin{figure*}[!ht]
\centering{
           \includegraphics[width=0.75\textwidth,clip=,trim=0mm 25mm 0mm 15mm]{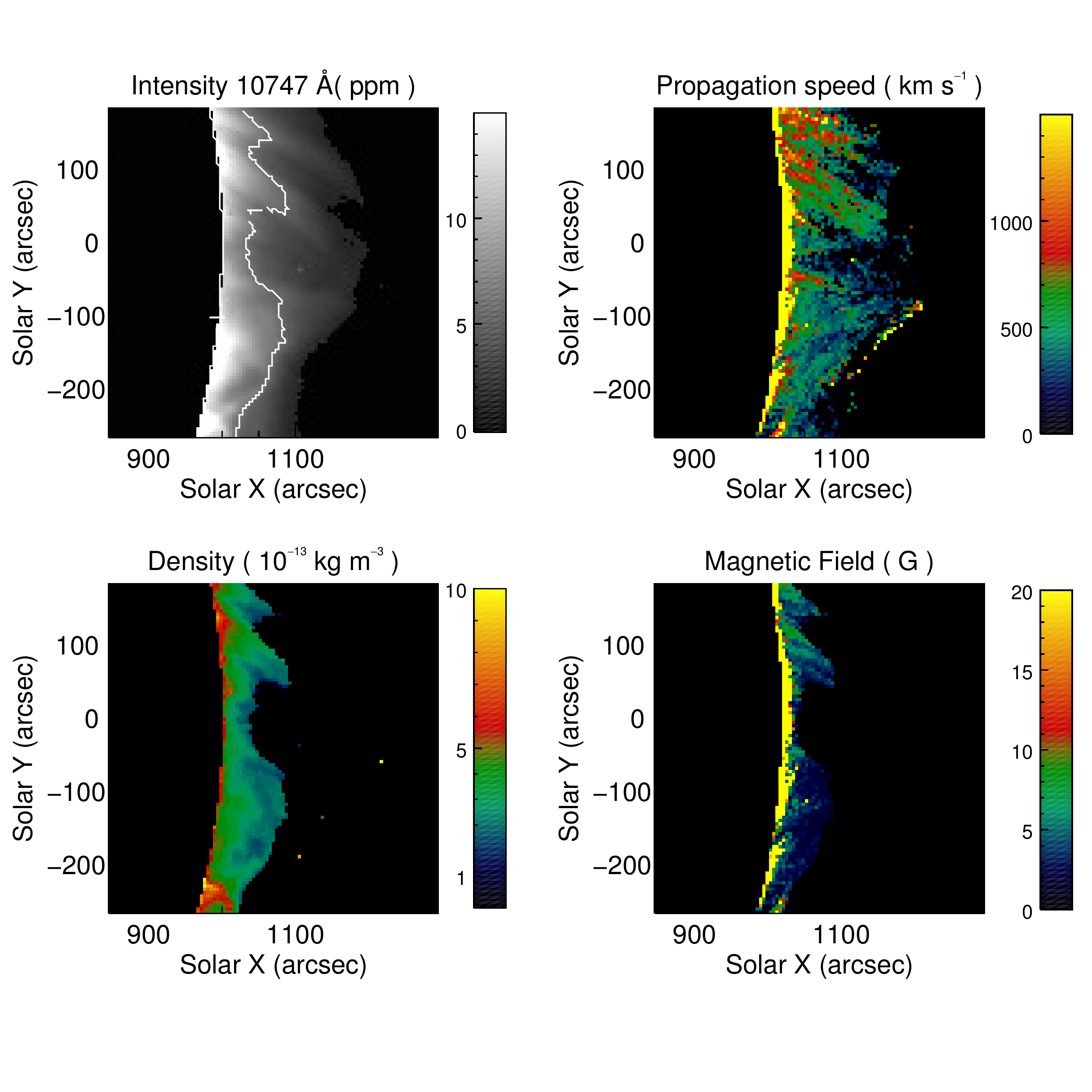}
           }
\caption{The results of the magneto-seismology of propagating transverse waves. The top left panel shows the coronal loop
	     as observed in 10747~{\AA} line centroid wavelength at 18:43~UT, with the measured propagation speeds of the transverse 
         waves shown in the top right panel. The white contour over the intensity image shows the region in 10798~{\AA} that has 
	     adequate signal level. The lower left panel displays the estimates of plasma density from the 10798~{\AA}/10747~{\AA} 
         ratio and the right hand panel shows the magnetic field estimate.}
\label{fig:ricks}
\end{figure*}

The EMD analysis suggests that the oscillation of the loop system is consistent with the fast magneto-acoustic kink mode 
as modelled by \citet{Aschwanden:1999}, with the clear Doppler velocity signal indicating a large-scale motion of the loop 
system towards and away from the observer. Figure~\ref{fig:imf_output} shows a damped oscillatory component present 
in both the line width and PBD intensity signals that has a period comparable to that observed in the Doppler velocity signal. 
However, it is very small and is most likely due to the apparent brightening and dimming as the loop system moves 
towards and away from the observer \citep[\cf.][]{Vandoor:2008b}. This interpretation also means that it may be possible to 
use the oscillation of the loop system to estimate its magnetic field via coronal seismology (Section~\ref{ss:seis}).

Although CoMP was originally designed to measure the coronal magnetic field, the seeing was not good enough for this 
event to make measurements of the full Stokes-\emph{I}, \emph{Q}, \emph{U} and \emph{V} parameters. Instead, a 
magnetoseismology technique developed by \citet{Tomczyk:2007} and \citet{Morton:2015,Morton:2016} was used to estimate the 
magnetic field strength of the loop system (Section~\ref{ss:comp}). This approach does not require any oscillation of 
the loop system, allowing an independent verification of the values derived from the coronal seismology technique. To provide 
an additional independent verification, the magnetic field was also estimated using a pair of magnetic field extrapolations 
derived from both GONG and HMI magnetograms (Section~\ref{ss:pfss}). 

\subsection{Coronal Seismology}\label{ss:seis}

The coronal seismology approach most commonly used to estimate the magnetic field within an oscillating coronal loop uses the 
damping of the loop as it returns to its original rest position to derive the period of oscillation. This is done by fitting 
the damped oscillation using an exponentially decreasing cosine function of the form,
\begin{equation}\label{eqn:cos}
d(t) = x_{0}~\mathrm{cos}\left(\frac{2 \pi t}{P} + \varphi \right)\mathrm{exp}\left(-\frac{t}{\tau}\right) + d_{0},
\end{equation}
where $x_{0}$ is the amplitude, $P$ is the period, $\varphi$ is the phase, $\tau$ is the damping time and $d_{0}$ 
is the equilibrium position. This model was applied to the variation in Doppler velocity shown in panel~b of 
Figure~\ref{fig:193_aia_comp}, with the resulting fit shown by the blue line and the fitted values given in the bottom 
right of the panel. It is clear that the model fits the observations very well, indicating that the assumption of a kink mode 
oscillation is consistent with the observations. The model was then applied to the Doppler velocity along the loop system, 
with Figure~\ref{fig:comp_dopp_profs} showing a representative sample of profiles with the blue line in each case showing 
the model fit to the data. It is clear that the oscillation is exhibited along the loop system at a range of heights and 
locations, with the model providing an excellent fit to the data in each case. 

Although the oscillation of the loop system may be interpreted as as a kink mode wave as discussed in Section~\ref{ss:emd}, 
it is not oscillating at the fundamental frequency. Instead, the out of phase Doppler signal observed between the legs of the 
loop system (apparent between $\approx$90-100$^{\circ}$ in the Doppler velocity plots shown in Figures~\ref{fig:193_aia_comp} 
and \ref{fig:comp_dopp_profs}) and the lack of a signal at the top of the loop system suggest that it is oscillating at the 2nd 
harmonic frequency. The oscillation period can therefore be used to estimate the strength of the magnetic field of the loop using 
the equation,
\begin{equation}
B = \frac{L}{P}\sqrt{2 \pi \rho_{in}\left(1+\frac{\rho_{in}}{\rho_{ex}}\right)}\label{eqn:bfield}
\end{equation}
where $B$ is the magnetic field strength, $L$ is the loop length, $P$ is the period of the oscillation, $\rho_{in}$ is the 
internal density of the loop system and $\rho_{ex}$ is the external density of the surrounding corona 
\citep[\cf.][]{Roberts:1984,Naka:2001,Aschwanden:2011}. 

The length of the loop was estimated by fitting an ellipse to the loop identified by visual inspection in the \emph{SDO}/AIA 
observations. This was done using the 193~\AA\ passband, with the images processed using the Multiscale Gaussian Normalisation 
technique of \citet{Morgan:2014} to highlight the loop and make it easier to identify. The process was repeated ten times to 
reduce uncertainty, with the loop length estimated at 711$\pm$8~Mm. 

The density of the loop system and the surrounding corona were estimated using the regularised inversion technique of 
\citet{Hannah:2013} assuming a temperature of $\sim$1.5~MK (corresponding to the peak emission temperature of the 193~\AA\ 
passband used to identify the wave--pulse). \corr{The spatial extent of the loop system was estimated using observations from the 
\emph{STEREO}-A spacecraft to be $\approx$373~Mm. This was used as the line-of-sight along which to integrate the emission 
measure for both the internal and external densities of the loop system.}
The internal density of the loop system at the location used to identify the oscillation in Doppler velocity was found using 
the emission measure at that location, with a mean internal density of \corr{$1.1\pm0.5 \times 10^8$~cm$^{-3}$} 
found across the loop system. In contrast, the external density was estimated using the mean density value for a 10 degree wide 
region of quiet Sun centered at 45 degrees clockwise from solar north at a comparable height to the measurement being made. This 
is indicated by the arc sector shown in Figure~\ref{fig:context}, and returned a mean external density of 
\corr{$2.5\pm2.1 \times 10^7$~cm$^{-3}$}
across the heights studied here.

Equation~\ref{eqn:bfield} was then used to estimate the magnetic field strength for a \corr{representative} range of \corr{five} 
oscillation periods and corresponding internal and external densities estimated along the loop. This returned an estimated 
magnetic field strength of \corr{$\approx5.5\pm1.5$~G}
within the loop system. While this is comparable to previous estimates using coronal 
seismology \citep[\eg,][]{Naka:2001}, suggesting that the approach is valid, it is quite low. There are most likely several 
reasons for this discrepancy. The oscillation observed here is the 2nd harmonic rather than the fundamental mode as is typically 
observed \citep[\eg,][]{Naka:2001,Vandoor:2008a,Aschwanden:2011,Verwichte:2013}. In addition, while every attempt has been made 
to minimise the uncertainty associated with this measurement, the large-scale nature of the loop system and the diffuse nature 
of the wave-pulse suggest that it is most likely a minimum uncertainty estimate.

\subsection{Direct measurement using CoMP}\label{ss:comp}

The magnetic field of the coronal loop can also be estimated from observations using magnetoseismology of the propagating 
transverse waves previously seen to be ubiquitous in CoMP Doppler velocities 
\citep[\eg,][]{Tomczyk:2007,Morton:2015,Morton:2016}. This offers an independent approach to estimate the magnetic field 
strength within the loop system. Following the approach developed by \citet{Tomczyk:2009,Morton:2015}, a coherence based 
method was used to track velocity perturbations in the Doppler velocities obtained from the 3-point 10747~\AA\ observations 
taken between 20:24:14--21:30:14~UT. This allowed both the direction and speed of the propagation to be measured as shown 
in the top right panel of Figure~\ref{fig:ricks}. 

The density of the loop system was estimated using the line centroid wavelength intensities from the 5-point 10747~\AA\ and 
10798~\AA\ data taken in the period 18:43:51--19:52:42~UT. This line pair is density sensitive \citep[\eg,][]{Flower:1973} and 
allows an estimate of the coronal density to be made for 12 pairs of 10747~\AA\ and 10798~\AA\ images \citep[CHIANTI V7.0;][]{Landi:2012} 
using the methodology outlined in \citet{Morton:2015,Morton:2016}. It was assumed that the variability of the density estimates 
between image pairs gives an reasonable estimate of the uncertainty associated with the density, not including any systematic 
errors associated with the uncertainties in atomic physics. The relative uncertainty in density is typically on the order of $10\%$, 
although it reaches $30\%$ in regions away from the loop of interest. Note that the observed emission from 10798~\AA\ is 
weaker than that from 10747~\AA, and the signal to noise ratio rises at a much greater rate as a function of height in the 
corona. The region of high quality signal in 10798~\AA\ is delimited by a white contour on the 10747~\AA\ image in the top left 
panel of Figure~\ref{fig:ricks} and the estimated plasma density is shown in the bottom left panel of Figure~\ref{fig:ricks}.

\begin{figure*}[!t]
\centering{
               \includegraphics[width=0.42\textwidth,clip=,trim=275mm 15mm 15mm 15mm]{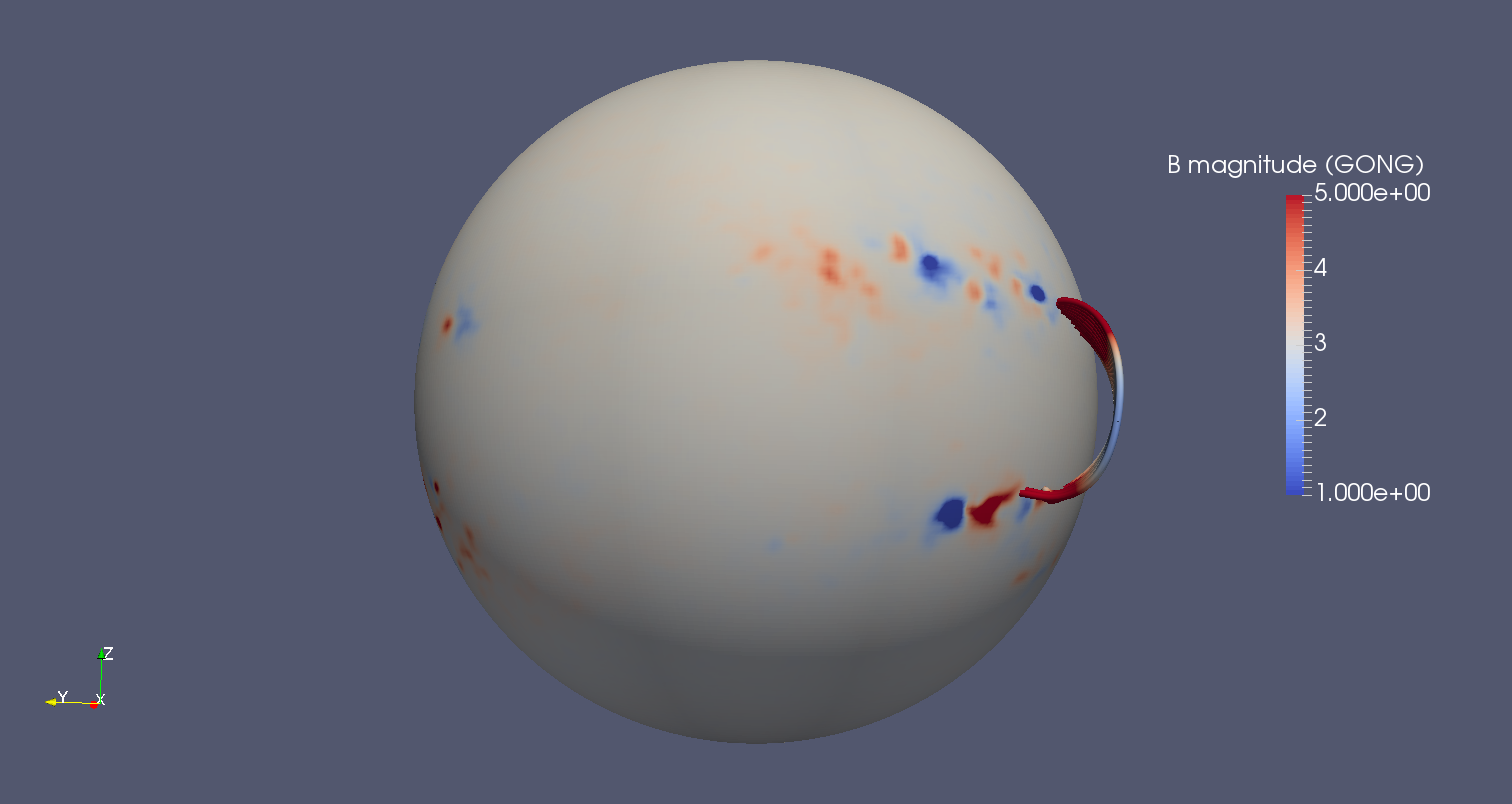}
               \includegraphics[width=0.42\textwidth,clip=,trim=275mm 15mm 15mm 15mm]{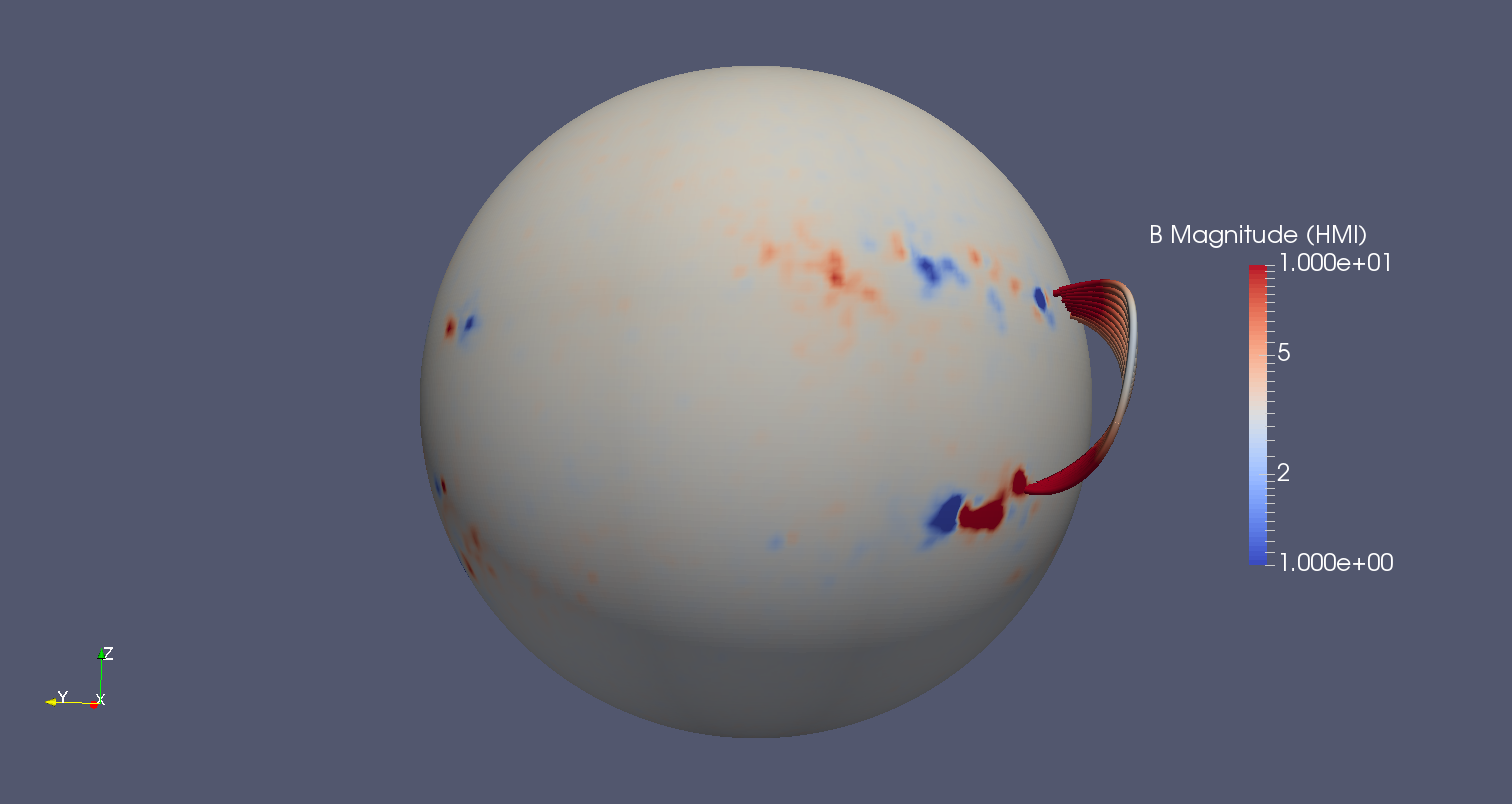}
               }
\caption{Potential Field Source Surface (PFSS) extrapolations using magnetograms from GONG (left panel) and \emph{SDO}/HMI 
		 (right panel). The line-of-sight averaged magnetic field along the line used in Figure~\ref{fig:193_aia_comp}
         yields $\sim$3.5~G for the GONG and $\sim$8~G for the \emph{SDO}/HMI magnetograms, respectively. \corr{Note that
         the colour bar in each panel indicates the variation in magnetic field strength of the corresponding loop system.}}
\label{fig:pfss_extrap}
\end{figure*}

The phase speed of the observed transverse waves is given by the kink speed,
$$
c_k^2=\frac{B_i^2+B_e^2}{\mu_0(\rho_i+\rho_e)},
$$ 
where \corr{$\mu_0$ is the permeability of a vacuum and} $i$ and $e$ refer to loop and ambient plasma values respectively 
\citep[\eg,][]{Nakariakov:2005}. Assuming that $B_i=B_e$ and taking the average density $\langle\rho\rangle$, an estimate for 
the magnetic field is given by,
$$
B=c_k\sqrt{\mu_0\langle\rho\rangle}.
$$
The average density is used as it reflects the fact that many oscillating structures are likely present within a single CoMP 
pixel \citep[typical coronal loop widths $\sim200-1000$~km, \eg,][]{Brooks:2013, Morton:2013}. As a result, both internal and 
ambient plasma will contribute to the observed emission. The estimated propagation speed and density may then be 
used to estimate the magnetic field, the results of which are displayed in the bottom right panel of Figure~\ref{fig:ricks}. The 
associated uncertainties are typically $\sim5\%$, reflecting the low errors associated with propagation speed determination.

\subsection{Magnetic field extrapolation}\label{ss:pfss}

A final independent estimate was made using the magnetic field extrapolated from photospheric magnetogram observations, 
comparable with previous approaches \citep[\eg,][]{Guo:2015}. In this case, two Potential Field Source Surface (PFSS) 
extrapolations were used to estimate the magnetic field in the solar corona corresponding to the loop system impacted by 
the pulse (see Figure~\ref{fig:pfss_extrap}). The first extrapolation shown in the left panel of Figure~\ref{fig:pfss_extrap} 
used a GONG magnetogram as a basis and estimated the global corona magnetic field using the finite differences method developed 
by \citet{Toth:2011}, also used in a study concerning the magnetic structure surrounding an AR \citep{Mandrini:2014}. The second 
extrapolation (right panel of Figure~\ref{fig:pfss_extrap}) was obtained from the PFSS package within SolarSoft described by 
\citet{Schrijver:2003} using a \emph{SDO}/HMI magnetogram as a basis. 

Figure~\ref{fig:pfss_extrap} shows that the two magnetograms initially used to extrapolate the coronal magnetic field are 
slightly different, as would be expected given the different sensitivity and resolution of the two instruments. This discrepancy 
affects the resulting extrapolated magnetic fields, so that the strength of the magnetic field along the transequatorial loop 
system is slightly different in both cases. The GONG (\emph{SDO}/HMI) extrapolations suggest a magnetic field of 
$\approx$5~G ($\approx$10~G) in the legs of the loop system, with the magnetic field in both cases dropping to $\approx$1~G 
at the loop-top. Along the line of sight used in Figure~\ref{fig:193_aia_comp}, the extrapolations return estimates of 
$\approx$3.5~G and $\approx$8~G for the GONG and \emph{SDO}/HMI magnetograms respectively. 

%

\section{Conclusion}\label{s:conc}

Here, we have used several independent techniques to estimate the magnetic field strength within a trans-equatorial loop system
following the impact of a global EUV wave-pulse. The initial impact of the wave-pulse drove a kink-mode oscillation of the loop 
system, allowing an estimate to be made of the magnetic field strength. This was then compared to the magnetic field strength 
obtained via magnetoseismology of the ubiquitous transverse waves previously observed by \citet{Tomczyk:2007} and 
\citet{Morton:2015,Morton:2016}. Finally, both sets of data-driven estimates were compared to extrapolated magnetic field 
measurements. All estimates were found to be broadly similar, consistent with previous results. \corr{This builds on work 
previously presented at the IAU Symposium on ``Solar and Stellar Flares and Their Effects on Planets'' \citep[described 
by][]{Long:2016} by using multiple independent techniques including direct measurement using CoMP and EMD analysis of the CoMP 
Doppler oscillation to determine the magnetic field strength of the loop system.}

While previous observations have shown coronal loop oscillations initially driven by the impact of a global EUV wave 
\citep[\eg,][]{Ballai:2011,Guo:2015}, this event is unique for several reasons. The trans-equatorial loop system was located
adjacent to the erupting active region, with the result that the global EUV wave-pulse was only observed to propagate southward 
away from the active region. The eruption was also observed by the CoMP instrument, making it one of only a handful of events 
observed by CoMP \citep[\cf,][]{Tian:2013}. The oscillation of the loop system was only apparent in the CoMP measurements of 
Doppler velocity, indicating that the erupting active region was closer to the observer than the loop system along the line-of-sight.

This oscillation of the trans-equatorial loop system in CoMP Doppler velocity was identified along the loop, and was measured 
by fitting an exponentially decaying cosine function. It was found that while the amplitude of the oscillation decreased with height, 
the period and damping time were found to be comparable along the loop. This suggests that the wave-pulse initially impacted 
the leg of the loop, which is consistent with the large-scale nature of the loop system. 

Despite the clear oscillation in CoMP Doppler velocity, no clear oscillation was observed in either the CoMP line width or the 
AIA intensity. However, oscillation was observed in both measurements when processed using an Empirical Mode 
Decomposition. This approach allowed an oscillation period and damping rate to be estimated from the line width and AIA intensity 
measurements, both of which are comparable to the CoMP Doppler velocity measurements, albeit with a slightly lower value. 
Although this suggests that the EMD approach may be beneficial for measuring oscillations in future observations, it should be 
noted that some component mixing was observed, which resulted in an overestimation of the oscillation frequency when 
fitted with a single frequency. Despite this, the signal in AIA intensity identified using the EMD analysis confirmed the kink 
mode nature of the oscillation.

The kink mode nature of the observed oscillation allowed an estimate to be made of the magnetic field strength within the loop 
system. The oscillation was measured across the loop at a range of heights and position angles, giving a mean magnetic field 
strength of \corr{$\approx5.5\pm1.5$~G} along the loop. This is comparable to the magnetic field strength of $\approx$1--9~G 
estimated using the independent magnetoseismology approach of \citet{Morton:2015,Morton:2016}. It is also comparable to the 
extrapolated magnetic field obtained from both HMI and GONG magnetograms, suggesting that the approach is valid, albeit 
within the limitations previously discussed by \citet{Verwichte:2013}.

In addition to the oscillation of the loop system previously described, the EMD analysis may also have allowed information on 
the pulse itself to be discerned which is inaccessible without a time-frequency analysis. Two clear signals were identified in 
the CoMP line width and AIA intensity: the oscillation of the loop system and a nonlinear signal. This nonlinear signal may 
correspond to the global EUV wave-pulse itself, in which case it is observational confirmation of the suggestion previously made 
by \citet{Long:2011a} that global ``EIT waves'' may be treated as a linear superposition of sinusoidal waves within a Gaussian 
envelope. Alternatively it may be a complex signal resulting from the interaction of the pulse and the loop system. As this set 
of observations is not optimal for discriminating between these possibilities, we intend to identify and further investigate the 
signal in future work.

\begin{acknowledgements}
The authors wish to thank Lidia van Driel-Gesztelyi, Marco Stangalini, Pascal D\'{e}moulin, Cristina Mandrini, Francesco 
Zuccarello and Steve Tomczyk for useful discussions as well as Christian Bethge and Hui Tian for discussions on density 
measurements with CoMP. \corr{The authors also wish to thank the referee for their comments which helped to improve the 
article.} DML and RJM are Early Career Fellows funded by the Leverhulme Trust. DPS was funded by the US Air Force Office 
of Scientific Research under Grant No. FA9550-14-1-0213. GV is funded by the Leverhulme Trust under Research Project Grant 
2014-051. \emph{SDO}/AIA data is courtesy of NASA/\emph{SDO} and the AIA science team. CoMP data is courtesy of the Mauna 
Loa Solar Observatory, operated by the High Altitude Observatory, as part of the National Center for Atmospheric Research 
(NCAR). NCAR is supported by the National Science Foundation.
\end{acknowledgements}


\end{document}